\documentstyle[preprint,aps]{revtex}

\renewcommand{\today}{23 April 1998}

\newcommand{\nc}{\newcommand}
\nc{\be}{\begin{equation}}
\nc{\ee}{\end{equation}}
\nc{\bea}{\begin{eqnarray}}
\nc{\eea}{\end{eqnarray}}
\nc{\beas}{\begin{eqnarray*}}
\nc{\eeas}{\end{eqnarray*}}
\nc{\noi}{\noindent}
\nc{\sD}{\not \! \! D}
\nc{\s}[1]{\not \! #1}
\nc{\non}{\nonumber}
\nc{\bb}{\bibitem}
\nc{\lf}{\left}
\nc{\ri}{\right}
\nc{\mb}[1]{\makebox[#1]{}}
\nc{\pa}{\partial}
\nc{\sA}{\not \! \! A}
\nc{\h}{\frac{1}{2}}
\nc{\ra}{\rightarrow}
\nc{\la}{\leftarrow}
\nc{\ep}{$e^+e^-\ra\pi^+\pi^-\;$}
\nc{\emuon}{$e^+e^-\ra\mu^+\mu^-\;$}
\nc{\epp}{$e^+e^-\ra\pi^+\pi^0\pi^-\;$}
\nc{\elec}{$e^+e^-\ra\gamma^*\ra e^+e^-\;$}
\def\mathunderaccent#1{\let\theaccent#1\mathpalette\putaccentunder}
\def\putaccentunder#1#2{\oalign{$#1#2$\crcr\hidewidth
\vbox to.2ex{\hbox{$#1\theaccent{}$}\vss}\hidewidth}}

\nc{\ti}{\mathunderaccent\tilde}
\nc{\M}{{\cal M}}
\nc{\rw}{$\rho\!-\!\omega\;$}
\nc{\bold}[1]{\mbox{\boldmath $#1$}}
\nc{\lrpa}{\stackrel{\leftrightarrow}{\pa}}

\begin{document}
\tightenlines
\draft
\preprint{\vbox{\null \hfill UK/TP 98--01  \\
                                       \null \hfill LPNHE 98--01 \\
                                       \null\hfill hep-ph/9804391}}
\title{SU(3) breaking and Hidden Local Symmetry\\
	{[Phys. Rev. D58, 074006 (1998)]} }
\author{M.~Benayoun}
\address{LPNHE des Universit\'es Paris VI et VII--IN2P3, Paris,
France\\
benayoun@in2p3.fr}
\author{H.B. O'Connell}
\address{Department of Physics and Astronomy, \\ University of Kentucky,
        Lexington, KY 40506-0055 USA\\
        hoc@pa.uky.edu }

\date{\today}
\maketitle

\begin{abstract}
We study the various existing implementations of SU(3) breaking
in the Hidden Local Symmetry model for the low energy hadronic sector
following a mechanism originally proposed by Bando, Kugo and Yamawaki (BKY).
We pay particular attention to hermiticity and current conservation.
Following this, we present a new method for including
symmetry breaking effects  which preserves the BKY mass relation
among vector mesons.
Symmetry breaking (SB) necessarily requires
a {transformation of the} pseudoscalar fields,
which, following BKY, we refer to as field renormalization.
We examine
the consequences of propagating this through all Lagrangian terms
including the anomalous ones.
We thus explore the consequences of these various SB schemes for
both charged and neutral pseudoscalar decay constants
as measured in weak and anomalous decays respectively.

{\it Keywords: Flavor symmetries, Chiral symmetries,
Currents and their properties, Chiral Lagrangians, Vector-meson dominance,
$\pi$, $K$, and $\eta$ mesons.}

\end{abstract}

\pacs{PACS numbers: 11.30.Hv, 11.30.Rd,
11.40.-q, 12.39.Fe,
12.40.Vv,
14.40.Aq}

\narrowtext
\section{Introduction}

In a recent paper \cite{ben4}, it has been shown that the pion form
factor is described in a perfectly coherent way by the Hidden Local
Symmetry model (HLS) proposed in Ref.~\cite{BKUYY}. Noticeably, it
has been shown, that a $\gamma \pi^+ \pi^-$ contact term is preferred
in fits with a coupling strength $c$ given by
\begin{equation}
c=\displaystyle 1 - \frac{f_{\rho \gamma}g_{\rho \pi \pi}}{m_{\rho}^2}
\label{bando}
\end{equation}
as predicted by the HLS model, where $ f_{\rho \gamma}$ and
$g_{\rho \pi \pi}$ are the usual coupling constants to
respectively the photon and a pion pair. In addition to
providing a nice description of the 
\ep data, it was also shown that the resulting phase of $F_{\pi}(s)$
accounted for the predicted $\pi \pi$ phase shift\cite{frog}
up to about 1 GeV/c, without further constraint. Moreover, the
values of $F_{\pi}(4 m_{\pi}^2)$ and for the $p$--wave scattering
length were found to agree perfectly with
chiral perturbation theory (ChPT) predictions
(see for instance Ref.~\cite{knecht}). This gives a hint that the
HLS model could successfully describe other scattering
data and that its extension to the anomalous
sector \cite{FKUTY} could describe radiative decays of light flavor mesons,
with a very small number of free parameters.

In the sector explored by Ref. \cite{ben4}, one does not expect effects of
the SU(3) symmetry breaking produced by the large mass difference
between the $s$ quark and the light $u$ and $d$ quarks.  Other sectors like
the kaon form factors or most radiative decays are surely more
sensitive to this. Therefore, a study
of symmetry breaking within the HLS model is an {\it a priori}
condition toward a full study of its relevance to low energy particle
physics in sectors mixing vector and pseudoscalar mesons explicitly.
In order to produce this breaking,
Bando, Kugo and Yamawaki (BKY)
\cite{BKY} proposed
a way which leaves the $u/d$ sector of pseudoscalar
mesons unchanged while sharply breaking the  $s$ sector. We will refer 
to this
(outlined in section {\bf 3}) as the BKY mechanism.

Much attention to this SU(3)
symmetry breaking has focussed
on its consequences for the
anomalous meson sector  \cite{hajuj,BGP,hash}
including proposed symmetry breaking variants \cite{HS}.
The aim of this paper is to study the consequences
of the original BKY mechanism
\cite{BKY} in both the anomalous and non-anomalous sectors.
We use the full
pseudoscalar field matrix ({\it i.e.} including the isoscalar sector),
in order to examine the hermitian version of the original BKY Lagrangian
\cite{BKY}, as well as one proposed by Bramon, Grau and Pancheri
(BGP) \cite{BGP} and one  we introduce here. Our
scheme allows one to recover interesting  properties of
both the BKY and BGP schemes, namely the mass relation among vector
mesons of Ref. \cite{BKY}, and
the current structure and conservation properties
obtained with the BGP scheme. In this
scheme the current coupling a vector meson
to a pseudoscalar pair, $P P'$, has a divergence
proportional to $m_P^2-m_{P'}^2$ which vanishes {for massless
$P$ and $P'$}. We shall refer to this throughout as current conservation,
because we only consider the case of massless pseudoscalar mesons.
Then in the physical case of massive pseudoscalar mesons current 
conservation
is broken in the appropriate way, {\it i.e.}
only by terms proportional to $m_P^2-m_{P'}^2$.

As recognized by BKY, their symmetry breaking
mechanism leads to a redefinition (we shall call this renormalization) of
the pseudoscalar fields, which has to be propagated to all Lagrangian
contributions. Focusing on the anomalous
(Wess--Zumino--Witten) WZW terms \cite{WZ,witten}, we show
that the symmetry breaking in the non-anomalous HLS Lagrangian,
produces {in this way} a new breaking of the anomalous terms and we 
illustrate
why it does not exhaust all breaking effects
(for example, {these} do not include loop effects
\cite{DHL}). This is of relevance
for the physics of $\eta/\eta'$ mesons, which has recently received
much interest from various points of view
\cite{ben1,ben2,ben3,VH,IM-BES-BG,FKS,FK,KP,Leutw}.

The paper is organized as follows, in section {\bf 2} we
briefly review the basics
of the HLS model. Section {\bf 3} is devoted to
an analysis of three ``natural''
variants of the original BKY mechanism for SU(3) symmetry breaking. We
show that, even when made hermitian, the BKY scheme  does not
separately conserve all currents occurring
in the interaction Lagrangian, {in the sense given above,
while the unbroken HLS Lagrangian does}.
The variant proposed by BGP
\cite{BGP} does, but gives the vector meson masses the standard
Gell-Mann--Okubo formula. We  propose another variant which
allows one to obtain the phenomenologically successful
BKY mass formula ($m_{\phi}m_{\omega}=m_{K^*}^2$)
and conservation
of all currents. We illustrate how the BKY mechanism, together
with a departure from ideal $\omega-\phi$ mixing,
generates a mass difference
$m_{\omega}-m_{\rho}$ which goes to zero with the symmetry breaking
parameter. In section {\bf 4}, we examine the consequences
of the pseudoscalar field renormalization implied by the BKY mechanism,
in (re--)deriving the decay constant $f_K$ and we show how
box and triangle anomalies are affected.
Most lengthy expressions are left to the appendix.

\section{Hidden Local Symmetry}

\indent \indent
We shall examine the low energy sector, including the octet and singlet
pseudoscalars within the context of the HLS model.
Here we present a brief {account} of the HLS
\cite{BKUYY,FKUTY,BKY} model.
The HLS model allows us to produce a theory with vector mesons as the gauge
bosons of a hidden local symmetry. These then become massive due to
the spontaneous breaking of a chiral $U(3)_L\otimes U(3)_R$
global symmetry. Let us
consider the chiral Lagrangian \cite{WCCWZ},
\be
{\cal L_{\rm chiral}}=\frac{1}{4}{\rm Tr}
[\pa_\mu  F\pa^\mu F^{\dagger}],
\label{ccwz}
\ee
where $F(x)=f_{P}U(x)$ in normal notation
and $f_{P}$ is a constant with dimensions of mass.
{In practice, one identifies this parameter with the
pion decay constant $f_P=f_{\pi}=93$ MeV}.
This exhibits the chiral $U(3)_L\otimes U(3)_R$ symmetry under
$U\ra g_LUg_R^{\dagger}$.
We can write this in exponential form and expand
\be
F(x)=f_{P} e^{2iP(x)/f_{P} }=f_{P} +2iP(x)-2P^2(x)/f_{P}+\cdots
\ee
therefore, substituting into Eq.(\ref{ccwz}) we see the vacuum corresponds
to $P=0$, $U=1$. That is, $F$ has a non-zero vacuum expectation
value which spontaneously breaks the $U(3)_L\otimes U(3)_R$
symmetry \cite{KR}. The massless Goldstone bosons contained in $P$,
then correspond to the perturbations
about the vacuum and we can think of expansions in this field
given by the hermitian matrix
$P=P^aT^a$ where the SU(3) generators
are normalized such that ${\rm Tr}[T^aT^b]=\delta^{ab}/2$.
Thus, for the pseudoscalars one has
\be
P=\frac{1}{\sqrt{2}}
  \left( \begin{array}{ccc}
            \frac{1}{\sqrt{2}}\pi^0+\frac{1}{\sqrt{6}}\pi_8+
            \frac{1}{\sqrt{3}}\eta_0&\pi^+ &  K^+ \\
            \pi^-  & -\frac{1}{\sqrt{2}}\pi^0+\frac{1}{\sqrt{6}}\pi_8
            +\frac{1}{\sqrt{3}}\eta_0  &  K^0 \\
            K^-             &  \overline{K}^0  &
             -\sqrt{\frac{2}{3}}\pi_8 +\frac{1}{\sqrt{3}}\eta_0 \\
         \end{array} \label{pseudoscalar}
  \right),
\ee
where we have included the singlet field $\eta_0$ assuming nonet
symmetry.{\footnote{
One could allow for departure from nonet symmetry by affecting the $\eta_0$
field by a multiplying parameter $x$ to be fixed by the data.
{Moreover, we will not address here
the problem of the renormalisation scale dependence
associated with the singlet pseudoscalar field (see \cite{KP,Leutw}
for instance).}
}}

However, in addition to the global chiral symmetry, QCD possesses a local
symmetry. The HLS scheme includes
such a symmetry in Eq.(\ref{ccwz}) in the following way. Let
\be
U(x)\equiv \xi^{\dagger}_L(x)\xi_R(x)\label{f-def}
\ee
where the $\xi$ fields undergo a local transformation, $h(x)$,
which does not affect the chiral field $U(x)$. In addition to
pseudoscalar fields, $P(x)$, the $\xi$ fields also possess a scalar
constituent $S(x)$, and are thus characterized by
\be
\xi_{R,L}(x)=e^{iS(x)/f_S}e^{\pm iP(x)/f_{P}},\mbox{               }
\xi_{R,L}(x)\ra h(x)\xi_{R,L}(x)g^{\dagger}_{L,R}.
\ee
As can be seen from Eq.~(\ref{f-def}), ${\cal L_{\rm chiral}}$ is
obviously invariant under this
local transformation.
{From} now on, as per usual \cite{BKUYY},
we remove $S(x)$ and thus $\xi_L^{\dagger}=\xi_R=\xi$.
We may rewrite ${\cal L_{\rm chiral}}$
explicitly in terms of the $\xi$ components
\be
{\cal L_{\rm chiral}}=-\frac{f_{P}^2}{4}{\rm Tr}\left[
(\pa_\mu\xi_L\xi_L^{\dagger}-\pa_\mu\xi_R\xi_R^{\dagger})\right]^2
\label{doon}
\ee
The Lagrangian can be gauged for both electromagnetism and the
hidden local symmetry by changing to covariant derivatives
\be
D_\mu\xi_{L,R}=\pa_\mu\xi_{L,R} -ig{\cal V}_\mu\xi_{L,R}+ie\xi_{L,R}A_\mu Q
\ee
where $A_\mu$ is the photon four-vector and $Q=$diag$(2/3,-1/3,-1/3)$
the charge matrix. The vector field,
${\cal V}={\cal V}^aT^a$, transforming
locally  as ${\cal V}_\mu\longrightarrow
h(x){\cal V}_\mu h^{\dagger}(x)+{i}h(x)
\pa_\mu h^{\dagger}(x)/{g}$, is given by
\be
{\cal V}=\frac{1}{\sqrt{2}}
  \left( \begin{array}{ccc}
   (\rho^0+\omega)/\sqrt{2}  & \rho^+             &  K^{*+} \\
            \rho^-           & (-\rho^0+\omega)/\sqrt{2}    &  K^{*0} \\
            K^{*-}           & \overline{K}^{*0}  &  \phi   \\
         \end{array}\label{vector}
  \right).
\ee
In Eq.~(\ref{vector})
$\omega$ and $\phi$ correspond to the ideally mixed states.
The HLS Lagrangian is then given by
${\cal L}_{\rm HLS}={\cal L}_{A}+a {\cal L}_V$ where
\bea\non
{\cal L}_{A}&=&-\frac{f_{P}^2}{4}{\rm Tr}\left[
D_\mu\xi_L\xi_L^{\dagger}-D_\mu\xi_R\xi_R^{\dagger}\right]^2
\equiv-\frac{f_{P}^2}{4}{\rm Tr}[L-R]^2
\\
{\cal L}_V&=&-\frac{f_{P}^2}{4}{\rm Tr}\left[
D_\mu\xi_L\xi_L^{\dagger}+D_\mu\xi_R\xi_R^{\dagger}\right]^2
\equiv-\frac{f_{P}^2}{4}{\rm Tr}[L+R]^2
\label{mass1}
\eea
and $a$ is a parameter which is not fixed by the theory.
However, setting $a=2$ allows one to recover the usual expression for
vector meson dominance (VMD) \cite{BKUYY} and moreover, there
is some experimental evidence \cite{ben4} that $a$
is slightly (but significantly) greater than 2. For this reason we shall
keep track of $a$ in the following expressions.

The full HLS Lagrangian  is somewhat lengthy, so we leave it to
the appendix, where it is given by
Eq.~(\ref{lag1}).
The photon and vector mesons acquire Lagrangian masses through
an analogue of the Higgs-Kibble mechanism; {we will refer
to these masses as HK masses.}\footnote{In light of
this model for low energy QCD
it is interesting to note the electric--magnetic
duality where the elementary electric degrees of freedom
become strongly coupled, leading to confinement. The magnetic degrees of
freedom, which are in the Higgs phase,
can be described as composites of the electric ones. These
magnetic particles
typically include massless gauge bosons associated with a new magnetic
gauge symmetry not present in the fundamental electric theory \cite{SS}.}
The photon, though, is seen to be massless
once the vector meson corrections to the vacuum polarization are
included, thus preserving EM gauge invariance (for a
fuller discussion of
this point see, for example, Ref.~\cite{review}).
One should also notice that the singlet field $\eta_0$ does not appear
in the SU(3) symmetric HLS Lagrangian.

\section{Flavor Symmetry Breaking}

\indent \indent To account for
deviations from SU(3) flavor symmetry in the low energy
sector Bando, Kugo and Yamawaki (BKY) \cite{BKY} introduced
symmetry breaking terms to the HLS Lagrangian,
as $(3,3^*)+(3^*,3)$ representations. However, there is no unique way
to do this, {though naturally one has to recover the unbroken case
smoothly when the symmetry breaking parameter goes to zero.}
The initial BKY symmetry breaking scheme
was recognized as
being non-Hermitian by BGP who
proposed a variant which restores Hermiticity \cite{BGP}.
We shall now discuss each variant of the original BKY symmetry breaking
scheme \cite{BKY} in detail {and present a new one.}

\subsection{The BKY Scheme}

The BKY scheme introduces the symmetry breaking
term $\epsilon={\rm diag}(0,0,c)$ through
\be
{\cal L}^{\rm BKY}_{{A}{(V)}}=-\frac{1}{4}f_P^2{\rm Tr}
[(D_\mu\xi_L\xi^{\dag}_L+D_\mu\xi_L\epsilon_{{A}({V})}\xi_R^{\dag})
\mp(D_\mu\xi_R\xi^{\dag}_R+D_\mu\xi_R\epsilon_{{A}({V})}\xi_L^{\dag})]^2
\ee
where the subscripts {$A$ and $V$ respectively
correspond to the $-$ and $+$ signs in the RHS of this expression.}
The relevant components are, defining $X_{A,V}=
(1+\epsilon_{A,V})$
\bea\non
{\cal L}^{\rm BKY}_{A}\!\!&=&\!\!{\rm Tr}[(\pa P X_A )^2
-i(g{\cal V}(P\epsilon_A+\epsilon_AP)-e(PA-AP+PA\epsilon_A+A\epsilon_AP))
\pa PX_A\\
\non
&&\hspace{2.15cm}+\:i(g(P\epsilon_A+\epsilon_AP){\cal V}
-e(AP-PA+P\epsilon_A A+\epsilon_A AP))X_A\pa P]
\\    \non
{\cal L}^{\rm BKY}_{V}\!\!&=&\!\!{\rm Tr}[
f_P^2(g{\cal V}X_V-eAX_V)^2+2i(g{\cal V}-eA)X_V\pa P(1-\epsilon_V)P
\\&~&+i(eA-g{\cal V})X_V(\pa PP+P\pa P)X_V]
\label{BKY1}
\eea
where we have assumed the appropriate contractions over the Lorentz indices
of ${\cal V}_\mu$, $A_\mu$ and $\pa_\mu$.
Eq.~(\ref{BKY1}) can easily be made Hermitian through the redefinition
\bea
{\cal L}_{\rm BKY}\longrightarrow\frac{1}{2}\left(
{\cal L}_{\rm BKY}+{\cal L}_{\rm BKY}^{\dagger}
\right)
\eea
where one recovers
smoothly the unbroken Lagrangian in the limit $X_{A(V)} \rightarrow 1$, as
desired.
This hermitian version of the BKY Lagrangian is given in
Eq.~(\ref{BKY2}).
One should note {here} that the BKY implementation of 
flavor symmetry breaking
produces an interplay of the singlet $\eta_0$ field which is absent
in the unbroken Lagrangian.
{From} Eq.~(\ref{BKY2}) we see
the BKY relation for the vector meson masses
\be
\frac{m_{K^*}}{m_\omega}=\frac{m_\phi}{m_{K^*}}=\sqrt{1+c_V}.
\label{BKY-mass}
\ee
This is very well fulfilled by the Breit--Wigner (BW) masses
of the corresponding vector mesons \cite{PDG} for $c_V\sim 0.3$.
Whether it should
also be true for the HK masses is presently an open
question{\footnote{The
vector meson masses reported in the Review of Particle Properties
are quite generally obtained from parametrizations assuming the 
(Breit--Wigner)
form $s-m_V^2-im_V \Gamma_V(s)$ for the vector meson propagators 
in fitting expressions. In order
to get an (approximate) estimate of the HK masses, one should rather use
propagators written like $s-m_V^2-\Pi_V(s)$, where $\Pi_V(s)$ is 
the vector
meson vacuum polarization. Although  there
is some hint  \cite{KKW}, that the meson masses
defined in this way could be significantly different from the usual
(BW) masses,  other studies predict negligible contributions
{from the real part of pseudoscalar meson loops}
to vector meson masses, apart from $\rho\ra2\pi\ra\rho$ \cite{craig}.
This is further complicated by the model dependence of traditional
mass extractions \cite{GO} and
we shall not discuss this matter any further here.}}.
 One should also note that no mass splitting is generated among
the $\rho$ and $\omega$ mesons.

In QCD the divergence of a general vector current
$J_\mu=\bar{a}\gamma_\mu b$ is proportional
to the quark mass difference $(m_a-m_b)$. In ChPT
\cite{GL465} the squares of
the pseudoscalar masses are proportional to linear combinations of the
quark masses, and hence to maintain this connection current divergences
should be of the form $(M_A^2-M_B^2)$ and so {should vanish} 
in the absence of
pseudoscalar meson mass terms. Therefore
 it is reasonable to use current conservation {(as defined above)}
to constrain parameters and symmetry breaking mechanisms
in the interaction Lagrangian.\footnote{
{From} a phenomenological point of view,
this assumption ensures that the coupling of a vector meson to 
two pseudoscalars
can be generally written 
$\epsilon^{\lambda}_{\mu} \cdot(p_1 - p_2)^{\mu}$,
with the massive vector particle having three
polarization states (denoted by $\lambda$), as usual.
The assumption about current conservation
prevents to rather have $A \epsilon^{\lambda}_{\mu} \cdot(p_1 - p_2)^{\mu}
+B\epsilon^{\lambda}_{\mu} \cdot(p_1 + p_2)^{\mu}$.}
Eq.~(\ref{BKY2}) then leads {first} to the condition
\be
c_A=ac_V
\ee
where $a=2$ reproduces VMD \cite{BKUYY}. Usually, it is
simply {\em assumed} that $c_A=c_V$ \cite{BKY,hajuj}.
Examining the kaon form-factors at $s=0$ we find
\be
F_{K^+}(0)=1+c_A,\,\,\,\,\,\,\,F_{K^0}(0)=0.\label{check}
\ee

It is {thus } clear that
field renormalization is required and BKY \cite{BKY}
remarked that the appropriate field renormalization is
\be
P_{R}=({1+\epsilon_A})^{1/2}P({1+\epsilon_A})^{1/2}.
\label{renorm}
\ee
Indeed, in addition to normalizing the kaon charge (Eq.~(\ref{check}))
the pseudoscalar kinetic term is  restored to its canonical form
\be
{\cal L}^{\rm kinetic}=\pa K_R^-\pa K_R^++\pa K_R^0\pa\bar{K_R}^0
+\pa\pi_R^+\pa\pi_R^-+\frac{1}{2}
\left(\pa\pi_R^0\pa\pi_R^0+\pa\pi_R^8\pa\pi_R^8
+\pa\eta_R^0\pa\eta_R^0\right),
\ee
where the subscript $R$, stands for ``renormalized."
Unfortunately, this does not quite restore current conservation
in the  $K^*$ interactions terms as
is clear from Eq.~(\ref{BKY3}), which expresses
${\cal L}^{\rm int}(K^*,K,\pi^8,\eta_0)$, due to terms quadratic
in the symmetry breaking parameters.
There are two cases, where current conservation
({in the sense defined just above}) can be restored,
{as in the unbroken Lagrangian.}
The first one
is the unphysical case where $\eta_0=\sqrt{2}\pi^8$, the other
case is if $a=1$, the Georgi vector limit \cite{georgi}.
Pure VMD supposes
$a=2$, while existing data prefer a slightly larger value \cite{ben4}
$a \simeq 2.4$, inconsistent with $a=1$ anyway.
Stated otherwise, the original BKY
scheme, even if modified in order to restore hermiticity, does not
completely maintain current conservation {under physically acceptable
conditions.}

\subsection{The BGP Scheme}

Having noted the non--Hermiticity of the original BKY Lagrangian,
BGP \cite{BGP} proposed the following variant
to the original BKY scheme
\be
{\cal L}^{\rm BGP}_{A,V}=-\frac{f_{P}^2}{4}{\rm Tr}\:[
(D_\mu\xi_L\xi_L^{\dagger}\mp D_\mu\xi_R\xi_R^{\dagger})^2
(1+\xi_L\epsilon_{A,V}\xi_R^{\dagger}+\xi_R\epsilon_{A,V}\xi_L^{\dagger})]
\ee
yielding upon the relevant weak field expansion
\bea
{\cal L}^{\rm BGP}_{A}&=&{\rm Tr}[(\pa P\pa P-ie(\pa P(AP-PA)
+(AP-PA)\pa P))(1+2\epsilon_A)]\\
{\cal L}^{\rm BGP}_{V}&=&{\rm Tr}[(i/2\{[\pa P,P],(g{\cal V}-eA)\}+
egf_P^2\{{\cal V},A\}+f_P^2 g^2{\cal V}^2)(1+2\epsilon_V)].
\eea
The full BGP Lagrangian is given by Eq.~(\ref{BGP2}).
A check of the $K^*$ interaction terms finds current conservation
guaranteed, but as there is no connection between $c_A$ and $c_V$, they
remain independent parameters, {at this stage}.
One should note that the BGP variant
does not invoke the interplay of the singlet field $\eta_0$, as opposed
to the original BKY scheme which does.

The other results of the BGP symmetry breaking scheme were alluded to in
general by BKY for any Lagrangian lacking $\epsilon^2$ terms, and the
following relations are easily recognized as linear truncations of the
BKY results. The vector meson masses are given by the
Gell-Mann--Okubo formula,
\be
m_{K^*}^2-m_\omega^2=m_\phi^2-m_{K^*}^2=c_Vaf_P^2g^2
\ee
which is less phenomenologically successful than the BKY relation
and $m_{\omega}=m_{\rho}$. Using the numerical values
from PDG \cite{PDG}, this relation implies $c_V=0.3 \sim 0.4$.

It is clear from the expression of the
pseudoscalar kinetic energy term that the BKY prescription for field
renormalization cannot change it to the canonical form. In order
to get this, one has to perform another change of fields. As per usual,
we find  $\pi_R=\pi$ and $K_R=K/\sqrt{1+c_A}$,
where the subscript $R$ stands for ``renormalized fields,'' while
for the isoscalars, we have
\bea\non
\displaystyle \pi^8&=&\displaystyle \frac{1}{3} \left[
\frac{2}{\sqrt{1+2c_A}}+1\right]\pi^8_R +
\frac{\sqrt{2}}{3} \left[ 1- \frac{1}{\sqrt{1+2c_A}}\right]\eta^0_R\\[0.5cm]
\displaystyle \eta^0 &=&\displaystyle \frac{\sqrt{2}}{3} \left[ 1-
\frac{1}{\sqrt{1+2c_A}}\right]\pi^8_R +
\frac{1}{3} \left[\frac{1}{\sqrt{1+2c_A}}+2\right]\eta^0_R
\label{chfield}
\eea
which has clearly a smooth limit for $c_A \rightarrow 0$.
Therefore, the BGP prescription, even if slightly
more complicated to renormalize than the BKY
scheme, allows one to recover all expectations.
We have already commented on the
mass formula, which cannot be presently considered as fully conclusive.

\subsection{An alternate scheme}

We have shown that the BKY {scheme} (once made Hermitian) fails
to preserve current conservation for the $(\pi^8,\eta_0)$ sector, whilst
the BGP scheme, though ensuring current conservation,
{seems} less successful in reproducing the observed accepted vector
meson mass splitting \cite{PDG}. We therefore introduce breaking in the
HLS Lagrangian in such a way that the desirable features of
both previous studies are reproduced, namely the BKY mass formula
and current conservation in all interactions.
We generalize Eq.~(\ref{mass1}) through
\be
{\cal L}_{A,V}=-\frac{f_{P}^2}{4}{\rm Tr}[(L\mp R)
(1+(\xi_L\epsilon_{A,V}\xi^{\dagger}_R+\xi_R\epsilon_{A,V}\xi^{\dagger}_L)/2)
]^2.\label{ours}
\ee
\noindent which has also a smooth unbroken limit. The terms we will be
interested in are then given by
\be
{\cal L}_{A,V}=-\frac{f_{P}^2}{4}{\rm Tr}[(L\mp R)X_{A,V}(L\mp R)X_{A,V}].
\ee
and hence
\bea
{\cal L}_A&=&{\rm Tr}[\pa PX_A\pa PX_A+2ie(PA-AP)X_A\pa PX_A]
\label{NS0}\\
{\cal L}_V&=&{\rm Tr}[f_{P}^2((g{\cal V}-eA)X_V)^2
+i(g{\cal V}-eA)X_V(\pa PP-P\pa P)X_V]
\label{NS1}
\eea

It is obvious from these expressions that the field renormalization
prescription of BKY \cite{BKY} is relevant in  this new scheme.
The kinetic pseudoscalar term is renormalized by the same
procedure as for BKY.
The full expression for the corresponding Lagrangian is
given in Eq.~(\ref{ourL}).
We also see the quadratic BKY relation between the $\omega$, $K^*$
and $\phi$ masses of Eq.~(\ref{BKY-mass}). What is more, like BGP,
current conservation is explicitly guaranteed.
{As for the BGP scheme, in contrast with that of BKY,
the pseudoscalar singlet field $\eta_0$, does not occur in the broken
Lagrangian.}
This could be inferred by looking at Eqs. (\ref{NS0}) and
(\ref{NS1}) for which symmetry breaking enters only through
the combinations $X_{A,V}$
unlike Eq. (\ref{BKY1}). In this new scheme, as for BGP,
$c_A$ and $c_V$ remain unrelated.
 As a final check we examine
the kaon form factors and find Eq.~(\ref{check}) holds for general
$a$.

\subsection{The $\rho$--$\omega$ Mass Splitting}

All symmetry breaking schemes outlined above, predict no mass splitting of the
$\rho$ and $\omega$ mesons (loop effects could be important here
\cite{KKW,craig}).  It should however be remarked that what have been called,
up to now, $\phi$ and $\omega$ are ideally mixed states, where $\omega$ is
purely non--strange and $\phi$ is purely strange.  There is however, strong
experimental evidence that, even if the mixing is close to ideal, it is not
exactly ideal.  Then the question arises as to whether a departure from ideal
mixing can (or should) be accounted for at the level of the Lagrangian itself
{and if the BKY symmetry breaking mechanism is able to contribute to $\rho$ --
$\omega$ mass splitting.}  As ideal mixing is not a fundamental symmetry, this
may not be actually considered as a symmetry breaking effect.

For $\omega$ and $\phi$ being considered as the ideally mixed states, we can
define, the physical states{\footnote{ In order to stay consistent with the
usual custom in the Effective Lagrangian community, we shall use the ideal
$\phi$ to be $+|s\overline{s}>$, while another usual custom \cite{ben2,PDG}
prefers $-|s\overline{s}>$, which allows
one to get these ideally mixed states
from the standard isovector singlet and octet states by a normal rotation
matrix, without any change of sign.  }} $\omega_P$ and $\phi_P$, by
\bea
\omega_P &=&\omega \cos{\delta}+\phi \sin{\delta}, \hspace{0.5cm}
\phi_P =\phi \cos{\delta} - \omega \sin{\delta}
\label{mix1}
\eea
where $\delta$ can be determined by the ratio of the measured
coupling constants \cite{ben3} $g_{\phi \pi^0 \gamma}$ and
$g_{\omega \pi^0 \gamma}$ which is found to correspond to $\tan{\delta}$
and gives about 3.25 degrees \cite{hash,ben2,benA}. The corresponding vector
mixing angle $\theta_V$ is therefore of the order 32 degrees, slightly
smaller than
its ideal value.

If one performs this change of variables in the unbroken HLS Lagrangian
(see Eq.~(\ref{lag1})),
the mass term is strictly conserved, and only couplings of the physical
$\omega_P$ and $\phi_P$ to  pseudoscalar mesons are changed by terms
of the order $\sin{\delta} \simeq 6\times10^{-2}$ or
higher. In the broken Lagrangians,
the situation looks slightly different.
The transformation generates a $\rho$--$\omega$ mass difference
\bea\non
({m_{\rho}^2-m_{\omega}^2})/{m_{\rho}^2}&=&
c_V(2+c_V)\sin^2{\delta} \hspace{0.5cm} {\rm BKY, ~~New ~Scheme}\\
&=&2c_V \sin^2{\delta} \hspace{1.8cm}{\rm BGP}
\label{mix2}
\eea
while leaving the $\phi_P$ mass modified, with respect to the ideal
$\phi$, by a negligible amount (a factor of the order $\cos{\delta})$. 
If one
considers likely values for $c_V$ ($\simeq 0.3$), 
this generates a $\rho$--$\omega$
mass splitting of about 2 to 3 MeV.

However, the original mass term for vector mesons in the broken Lagrangians
also generates a transition term from $\omega_P$ to $\phi_P$
\bea\non
-\omega_P \phi_P \sin{2\delta}&\times&
c_V(2+c_V) \hspace{0.5cm} {\rm BKY, ~~New ~Scheme}\\
&\times&2c_V \hspace{1.8cm} {\rm BGP}
\label{mix3}
\eea
which adds $\omega_P-\phi_P$ direct 
transitions to the usual $\gamma$ to vector
meson direct transitions. It should be noted that such a term 
(which vanishes in
the unbroken limiting case), is small (its coupling is of the order 1\%
of $m_{\rho}^2$), and probably inefficient because of the large mass
difference between these mesons. Moreover, it comes supplementing already
existing transition effects by means of the $K \overline{K}$ loop
effects. Whether such a term could be experimentally visible is thus
not obvious to answer.
It is however interesting to see that a very small admixture of
$s \overline{s}$ inside the $\omega$ is able to generate and explain
a small mass splitting between the $\rho$ and $\omega$ mesons {by means of
the BKY symmetry breaking mechanism, which vanishes with the symmetry
breaking parameter.}

\section{Pseudoscalar Decay Constants and Anomalies}

We are now in a position to determine the pseudoscalar decay constants,
and examine some consequences of the pseudoscalar field renormalization
implied by the BKY symmetry breaking mechanism.
We have of course to distinguish the case of $\pi^0$, $\eta$ and $\eta'$,
which proceed from the low energy anomalous Lagrangians \cite{WZ,witten},
from $\pi^\pm$, $K^\pm$, $K^0$ and $\overline{K}^0$ mesons,
which can be determined from the pseudoscalar meson coupling
to an axial vector field.

\subsection{Decay Constants from Non--Anomalous Sector}

The charged pseudoscalar decay constants are measured in weak decays
$P^\pm\ra\ell^\pm+\nu_\ell$
and $P^\pm\ra\ell^\pm+\nu_\ell\gamma$ \cite{PDG}. Therefore to examine
this in the HLS model we need to include axial vectors, ${\cal A}_\mu$,
through \cite{Meiss}
\be
D_\mu\xi_L=(\pa_\mu-ig{\cal V}_\mu+ig_A{\cal A}_\mu)\xi_L,\,\,\,\,\,
D_\mu\xi_R=(\pa_\mu-ig{\cal V}_\mu-ig_A{\cal A}_\mu)\xi_R.\label{axials}
\ee
The
pseudoscalar decay constants defined through (note unlike
the mini-review
of Suzuki in Ref.~\cite{PDG} we include $\sqrt{2}$)
\be
\langle0|{\cal A}_\mu|P({\bold{q}})\rangle=i\sqrt{2}f_Pq_\mu.\label{decay}
\ee
are determined from the
${\cal A}^\mu\pa_\mu P$ interaction (set $g_A=1$), which for
${\cal L}^{\rm BKY}$ and ${\cal L}^{\rm new}$ is given by
\be
{\cal L}^{\rm BKY,new}_{{\cal A}\pa P}=-2 f_{P}  {\rm Tr}[
{\cal A}(1+\epsilon_A)\pa P(1+\epsilon_A)]
\ee
while for ${\cal L}^{\rm BGP}$ one has
\be
{\cal L}^{\rm BGP}_{{\cal A}\pa P}
=-f_P{\rm Tr}[(\pa P{\cal A}+{\cal A}\pa P)(1+2\epsilon_A)].
\ee
Constructing axial currents of the appropriate quark flavor one finds
for the renormalized pion and kaon fields in all three models
\bea
{\cal L}_{{\cal A}\pa P}=
-\sqrt{2}f_{P}[(\pa_\mu\pi^-_R+\pa_\mu\pi^+_R){\cal A}^\mu(ud)
+\sqrt{1+c_A}(\pa_\mu K^+_R+\pa_\mu {K}^-_R){\cal A}^\mu(us)]
\label{PSdecays}
\eea
and hence
\be
f_{\pi^+}=f_{P},\,\,\,\,\,\,
f_{K}=({1+c_A})^{1/2}f_{\pi^+}\\
\ee
and we see that $f_P$ is just the usual pion decay constant $\sim$ 93 MeV.
For the BKY scheme this gives a prediction for the kaon decay constant
\be
f_K=({1+c_A})^{1/2}f_{\pi^+}=({1+2c_V})^{1/2}f_{\pi^+}\sim 1.26 f_{\pi^+}
\ee
which is in very good agreement with experiment \cite{PDG}. For
both our
{new scheme} and that of BGP
$c_A$ is a parameter to fit to data, thus using
the experimental result $f_K/f_{\pi^+}=1.22$ we find $c_A=0.49$.
{It is likely that $c_A$ can also be derived from scattering data
like the kaon form factors or radiative decays.}

\subsection{The Anomalous Sector}

\indent \indent
The neutral decay constants, however, are measured in the
anomalous processes $P^0\ra\gamma\gamma$. Therefore we
cannot obtain them in the previous manner but rather
from analyzing the  anomalous Lagrangians.
We shall not consider explicit
symmetry breaking terms in the anomalous action, but rather
propagate the pseudoscalar field renormalization
into the anomalous Lagrangian. As will be seen, this induces
symmetry breaking effects in a way not previously reported.
The field renormalization are given by Eq.~(\ref{renorm})
for the original BKY scheme and our proposed method,
and by Eq.~(\ref{chfield}) for the BGP variant \cite{BGP}.

It has been claimed \cite{hajuj} that the singlet and octet decay constants
can be determined by renormalizing the $\pi^8$ and $\eta^0$ fields with the
(square root of) the coefficients in front of $(\partial \pi^8)^2$ and
$(\partial \eta^0)^2$
in Eq.~(\ref{BKY2}), leaving aside the mixed term
$\partial \pi^8\partial \eta^0$. Even if, numerically, the results could
look interesting \cite{hajuj} compared with expectations \cite{GL465,DW},
the procedure seems questionable, as any reference
to the triangle anomaly, which controls the two--photon decays
of $\pi^0,  ~\eta$ and $\eta'$, is missing.

A convenient form of the Wess--Zumino anomalous action \cite{WZ} was
constructed by Witten \cite{witten}, and although this has been
generalized to
include vector mesons \cite{FKUTY,hash,HS} we shall {here} 
consider only the soft
limit in which they play no role. {However, the field renormalization
performed in the HLS Lagrangian has clearly to be propagated to all
possible (anomalous) terms.}
The two relevant interactions are $\gamma PPP$ and $\gamma\gamma P$.
For the first we have
\be
{\cal L}_{\gamma PPP}=\frac{ieN_c}{3\pi^2f_{P}^3}
\varepsilon^{\mu\nu\alpha\beta}
A_\mu{\rm Tr}[Q\pa_\nu P\pa_\alpha P\pa_\beta P]\label{APPP}
\ee
where our $f_{P}$ is half Witten's $F_\pi$ \cite{witten}.
Renormalizing the bare pseudoscalar fields through Eq.~(\ref{renorm})
or Eq.~(\ref{chfield}) we express Eq.~(\ref{APPP})
in terms of the physical pseudoscalar fields,
\be
{\cal L}_{\gamma\pi^+\pi^-P^0}=-\frac{ieN_c}{12\pi^2f_{P}^3}
\varepsilon^{\mu\nu\alpha\beta}A_\mu\left[{\pa_\nu\pi^0_R}+
\frac{1}{\sqrt{3}}{\pa_\nu\pi^8_R}+\sqrt{\frac{2}{3}}
{\pa_\nu\eta^0_R}\right]
\pa_\alpha\pi^+_R\pa_\beta\pi^-_R
\label{box1}
\ee
No symmetry breaking results from field renormalization
in any of the three implementations of symmetry breaking
{for all $\gamma \pi^+ \pi^- P^0$ vertices; things
are, of course, different in other sectors. }
This is
easily understood in terms of the underlying quark substructure for
such a process in which the $s$ quark responsible for symmetry
breaking cannot contribute.
More precisely, this follows from the fact that all variants
of the BKY symmetry breaking, leave invariant the combination
of $\pi^8$ and $\eta^0$ fields (or of $\eta$ and $\eta'$ fields)
which corresponds to the $u\overline{u}+d\overline{d}$ field
component. This is indeed a specific feature of
all implementations of symmetry breaking discussed here.

The anomalous Lagrangian for the decay $P\ra\gamma\gamma$, is given by
\cite{witten}
\bea
{\cal L}_{\gamma\gamma P}&=&-\frac{N_ce^2}{16\pi^2f_{P}}
\varepsilon^{\mu\nu\alpha\beta}F_{\mu\nu}F_{\alpha\beta}{\rm Tr}
\left[Q^2P\right].
\eea
In the original BKY scheme, as well as the new scheme, we find
\bea
{\cal L}_{\gamma\gamma P}^{\rm BKY,new}&=&-\frac{N_ce^2}{48\pi^2f_{P}}
\varepsilon^{\mu\nu\alpha\beta}F_{\mu\nu}F_{\alpha\beta}\left[
\frac{\pi^0_R}{2}+\frac{3+5c_A}{6\sqrt{3}(1+c_A)}\pi^8_R
+\frac{6+5c_A}{3\sqrt{6}(1+c_A)}{\eta^0_R}\right],
\label{triangle2}
\eea
using Eqs. (\ref{renorm}), while using Eqs. (\ref{chfield}),
the BGP variant gives
\be
{\cal L}_{\gamma\gamma P}^{\rm BGP}=-\frac{N_ce^2}{48\pi^2f_{P}} \displaystyle
\varepsilon^{\mu\nu\alpha\beta}F_{\mu\nu}F_{\alpha\beta}\left[
\frac{\pi^0_R}{2}+
\frac{5 \sqrt{1+2c_A}-2}{6\sqrt{3}\sqrt{1+2c_A}}\pi^8_R+
\frac{5\sqrt{1+2c_A}+1}{3\sqrt{6}\sqrt{1+2c_A}}{\eta^0_R}\right].
\label{triangle2b}
\ee

Therefore the BKY mechanism for U(3) breaking
leads to a new modification of the anomaly equations which leaves
the usual box anomaly terms $\gamma \pi^- \pi^+ (\pi^0/ \eta/\eta')$
and the coupling $\gamma \gamma  \pi^0$ unchanged,
while changing only slightly the triangle anomaly equations for
$\gamma \gamma  (\eta/\eta')$ (see below for numerical estimates).
This is in contrast with Kisselev
and Petrov \cite{KP} who find a stronger breaking affecting both
the anomalous triangle and box couplings of pseudoscalar mesons, keeping
them structurally unchanged. Leutwyler on the other hand predicts
a deep change in the structure of the triangle anomaly
matrix element \cite{Leutw}.
Feldman and Kroll \cite{FK}, also break deeply the structure of the
triangle anomaly.

Examining the full effect of the BKY symmetry breaking schemes, requires a
refitting of all data on
the box anomaly in order to stay consistent with the HLS
model and accurately test all its assumptions in the anomalous sector
\cite{FKUTY}.  This work, which goes far beyond the aim of this paper, is
presently under way \cite{benA}.

Eq. (\ref{box1}) clearly shows that the change of fields
required by the BKY breaking mechanism does not exhaust the expected
symmetry breaking effects as this would imply $f_0=f_8=f_{\pi}$. Therefore
higher order effects have to be accounted for {when using} Eq.~(\ref{box1});
they can formally be described by changing appropriately a factor of $f_P$ to
$f_{\pi}$, $f_8$ and $f_0$ depending on the (renormalized) field it
multiplies ($\pi^0_R, ~\pi^8_R, ~\eta^0_R$), while the two remaining
powers of $f_P$ have to be changed to $f_{\pi}$. Consistency then
implies the corresponding changes in the relations for the triangle
anomaly couplings (\ref{triangle2}) and (\ref{triangle2b}).

Then, the couplings occurring in the $\gamma \gamma P$ sector
are to be affected by weighting factors, relative to the unbroken case,  
which
result in
\bea\non
\displaystyle \frac{1}{f_{\pi}} \!\!&\rightarrow&\!\!
\frac{1}{f_{\pi}},\hspace{0.3cm}
\displaystyle \frac{1}{f_8} \rightarrow \frac{1.22}{f_8}, \hspace{0.3cm}
\displaystyle \frac{1}{f_0} \rightarrow \frac{0.94}{f_0},
\hspace{0.5cm} {\rm ~~~BKY, New ~Scheme} \\
\displaystyle \frac{1}{f_{\pi}} \!\!&\rightarrow&\!\!
\frac{1}{f_{\pi}}, \hspace{0.3cm}
\displaystyle \frac{1}{f_8} \rightarrow \frac{1.19}{f_8}, \hspace{0.3cm}
\displaystyle \frac{1}{f_0} \rightarrow \frac{0.72}{f_0},
\hspace{0.5cm}  {\rm ~~~BGP}
\label{wzbrk}
\eea
using $c_A=0.49$.

Using Eqs.~(\ref{box1}) and (\ref{triangle2}) (or (\ref{triangle2b})),
one can easily write down the matrix elements for
$\gamma \pi^+ \pi^- (\pi^0/\eta/\eta')$ and  
$\gamma \gamma(\pi^0/\eta/\eta')$
after introducing the physical $\eta$ and $\eta'$ fields through
\cite{DHL,VH,FKS,PDG}
\be
\eta\equiv\cos\theta\;\pi^8_R-\sin\theta\;\eta^0_R,\hspace{1.3cm}
\eta'\equiv\sin\theta\;\pi^8_R+\cos\theta\;\eta^0_R,\label{rotn}
\ee
which leaves the pseudoscalar kinetic energy term with its canonical form
(no $\eta-\eta'$ mixing is introduced) for each
variant of the BKY SU(3) breaking schemes. This breaks the anomaly
set of equations in an original way, {\it i.e.} the box anomaly
equations (involving $\gamma \pi^+ \pi^- P^0$) are strictly
unchanged, while triangle anomaly couplings undergo modest
symmetry breaking effects under realistic conditions. This is
indeed obtained by including the $s$ quark symmetry breaking while
keeping the $u$ and $d$ degenerate.
Of course, all  anomalous terms within the HLS approach \cite{FKUTY},
undergo breaking by this field renormalization.

The full physics consequences of this breaking mechanism is under
consideration
\cite{benA}.

\section{Conclusion}

We have shown that a few variants of the SU(3) breaking mechanism proposed by
BKY \cite{BKY}, allow one to maintain current conservation {(in the sense
defined in the introduction, namely that currents are {\it strictly} conserved
for massless pseudoscalar mesons)} in all sectors of the broken HLS Lagrangian.
The original BKY breaking scheme, once hermitized, achieves this symmetry
breaking with a single parameter, since $c_A=a c_V$, but current conservation
in the full isoscalar sector implies the unrealistic condition that $a=1$, in
contradiction with VMD ($a=2$) and data ($a=2.4$).  It remains to find places
where possible departures from the usual assumption of the current conservation
(as we defined it) can be tested, and which arises only from symmetry breaking
{effects} in the HLS Lagrangian.

However, the new scheme we propose, as well
as the BGP scheme, maintain this current conservation. They mainly
differ {from each other} by the mass relation among vector mesons once
symmetry breaking has occurred.
The broken Lagrangians in both cases
depend on two independent breaking parameters $c_A$
and $c_V$. Data from $e^+ e^-$ annihilations
could allow one to fix them with a consideration
of radiative decays of light mesons
to check their consistency, and thus the relevance of
the BKY symmetry breaking mechanism. Moreover, some work
is still needed in order to see on whether the  mass
relation for vector mesons (Gell-Mann--Okubo versus
Bando--Kugo--Yamawaki) has to be considered conclusive.

We have also shown that the field ``renormalization"
{(or transformation)} following from the
BKY symmetry breaking mechanism, modifies in a very mild way the
anomalous WZW Lagrangian terms. This contrasts with several other
proposed mechanisms.

A full phenomenological study of this symmetry breaking mechanism
may allow us to answer the question of the experimental relevance
of the HLS Lagrangian. The present hint, relying on several
studies, is optimistic.

\vspace{1.5cm}
\begin{center}
{\bf Acknowledgements}
\end{center}
We would like to thank M.~Hashimoto, V.A.~Petrov  and C.D.~Roberts 
for helpful correspondence.
We also thank  C.~Carimalo, K-F.~Liu and W.~Wilcox for
interesting discussions
and comments. HOC is supported by the
US Department of Energy under grant DE--FG02--96ER40989.

\appendix

\section{Lagrangian expressions}

Here we present expressions for the full Lagrangians. {For all Lagrangians,
especially the SU(3) broken ones, expressions are given in terms of
bare fields in the sense of the BKY \cite{BKY} pseudoscalar field
renormalization.}

Defining $(P\stackrel{\leftrightarrow}{\pa}P'\equiv P\pa P'-P'\pa P)$
the original (unbroken) HLS Lagrangian can be expanded as follows
\bea \non
{\cal L}_{\rm HLS}&=&\frac{2}{3}ae^2f_{P}^2A^2
+\frac{af_{P}^2g^2}{2}(\rho^2+\omega^2+\phi^2)
+af_{P}^2g^2(\rho^+\rho^-+{K}^{*+}K^{*-}+\bar{K}^{*0}K^{*0})\\ \non
&-&\!\!aef_{P}^2g\left[\rho+\frac{\omega}{3}-\frac{\sqrt{2}}{3}\phi\right]
\cdot A
+\frac{i}{2}
\left[a g\rho+e(2-a)A\right]\cdot[\pi^-\lrpa\pi^+]\\
\non
&+&\!\!\frac{i}{4}\left[ag(\rho+\omega-\sqrt{2}\phi )
+2e(2-a)A\right]\cdot[K^-\lrpa K^+]\\
\non
&+&\!\!
\frac{iag}{4}\left[\rho-\omega+\sqrt{2}\phi \right]\cdot
   \left[K^0\lrpa\overline{K}^0 \right]\\ \non
&+&\!\!\frac{iag}{2\sqrt{2}}\rho^+\left[K^-\lrpa K^0
+\sqrt{2}\pi^0\lrpa\pi^-\right]
+\frac{iag}{2\sqrt{2}}\rho^-\left[\bar{K}^0\lrpa K^+
+\sqrt{2}\pi^+\lrpa\pi^0
\right]\\ \non
&+&\!\!\frac{iag}{4}K^{*0}\left[\bar{K}^0\lrpa\pi^0
+\sqrt{2}\pi^+\lrpa K^-
+\sqrt{3}\pi^8\lrpa\bar{K}^0\right]
\\ \non
&+&\!\!\frac{iag}{4}\bar{K}^{*0}\left[\pi^0\lrpa{K}^0
+\sqrt{2}K^+\lrpa\pi^-
+\sqrt{3}K^0\lrpa\pi^8\right]
\\ \non
&+&\!\!\frac{iag}{4}{K}^{*-}\left[{K}^+\lrpa\pi^0
+\sqrt{2}K^0\lrpa\pi^+
+\sqrt{3}K^+\lrpa\pi^8
\right]
\\
&+&\!\!\frac{iag}{4}K^{*+}\left[\pi^0\lrpa K^-
+\sqrt{2}\pi^-\lrpa\bar{K}^0
+\sqrt{3}\pi^8\lrpa K^-\right] \non \\
\label{lag1}
\eea

 The BKY Lagrangian, SU(3) broken as prescribed in Ref. \cite{BKY} is
after hermitization
\bea\non
{\cal L}^{\rm Herm}_{\rm BKY}&=&(1+c_A)\pa K^-\pa K^++(1+c_A)
\pa K^0\pa \bar{K}^0
+\pa\pi^+\pa\pi^-+(1/2)\pa\pi^0\pa\pi^0\\
\non&+&\!\!
\frac{1}{2}\left(1+\frac{4}{3}c_A+\frac{2}{3}c_A^2\right)\pa\pi^8\pa\pi^8
+\frac{1}{2}\left(1+\frac{2}{3}c_A+\frac{1}{3}c_A^2\right)\pa\eta_0\pa\eta_0
\\
\non&-&\!\!
\frac{\sqrt{2}}{3}c_A(2+c_A)\pa\eta_0\pa\pi^8
\\
\non&+&\!\!
ae^2f_{P}^2A^2\left[\frac{2}{3}+\frac{2c_V}{9}+\frac{c_V^2}{9}\right]
+ag^2f_{P}^2(\rho^+\rho^-+ (1+c_V)(K^{*-}K^{*+}+\bar{K}^{*0}K^{*0}))\\
\non&+&\!\!
\frac{1}{2}ag^2f_{P}^2({\rho^0}^2+\omega^2+(1+c_V)^2\phi^2)
 -aegf^2A\left[\rho^0 +\frac{1}{3}\omega-\frac{\sqrt{2}}{3} (1+c_V)^2\phi
  \right]\\
  \non&+&\!\!
\frac{ie}{3}A\left[(3(1-a/2)+ac_V/2+(3c_A-c_A^2)/2)
K^-\stackrel{\leftrightarrow}{\pa} K^+
+(ac_V+c_A^2/2)K^0\stackrel{\leftrightarrow}{\pa}\bar{K}^0\right]\\
  \non&+&\!\!
  ie(1-a/2)A\pi^-\stackrel{\leftrightarrow}{\pa} \pi^+\\
\non&+&\!\!
\frac{ig}{4}\rho^0\left[(a(1-c_V)+c_A)(K^-\stackrel{\leftrightarrow}{\pa} K^+
+K^0\stackrel{\leftrightarrow}{\pa}\bar{K}^0)
+2a\pi^-\stackrel{\leftrightarrow}{\pa}\pi^+\right]\\
\non&+&\!\!
\frac{ig}{2\sqrt{2}}\rho^+\left[(a(1-c_V)+c_A)
K^-\stackrel{\leftrightarrow}{\pa} K^0
+\sqrt{2}a\pi^0\stackrel{\leftrightarrow}{\pa}\pi^-\right]\\
\non&-&\!\!
\frac{ig}{2\sqrt{2}}\rho^-\left[(a(1-c_V)+c_A)
K^+\stackrel{\leftrightarrow}{\pa} \bar{K}^0
+\sqrt{2}a\pi^0\stackrel{\leftrightarrow}{\pa}\pi^+\right]\\
\non&+&\!\!
\frac{ig}{4}\omega(a(1-c_V)+c_A)\left[K^-\stackrel{\leftrightarrow}{\pa} K^+
-K^0\stackrel{\leftrightarrow}{\pa}\bar{K}^0\right]\\
\non&-&\!\!
\frac{ig}{2\sqrt{2}}
\phi(a(1+c_V)+c_A+c_A^2)\left[K^-\stackrel{\leftrightarrow}{\pa} K^+
-K^0\stackrel{\leftrightarrow}{\pa}\bar{K}^0\right]\\
\non&+&\!\!
\frac{ig}{4}K^{*0}\left[(a+c_A)(\bar{K}^{0}\pa\pi^0-\sqrt{2}K^-\pa\pi^+)
-a(1+c_V)(\pi^0\pa\bar{K}^{0}-\sqrt{2}\pi^+\pa K^-)\right]\\
&-&\!\!
\frac{ig}{4}\bar{K}^{*0}\left[(a+c_A)({K}^{0}\pa\pi^0-\sqrt{2}K^+\pa\pi^-)
-a(1+c_V)(\pi^0\pa{K}^{0}-\sqrt{2}\pi^-\pa K^+)\right]\non \\
&+&\!\!
\frac{ig}{4}K^{*-}\left[(a+c_A)(K^+\pa\pi^0+\sqrt{2}K^0\pa\pi^+)
-a(1+c_V)(\pi^0\pa K^++\sqrt{2}\pi^+\pa K^0)\right]\non \\
&-&\!\!
\frac{ig}{4}K^{*+}\left[(a+c_A)(K^-\pa\pi^0+\sqrt{2}\bar{K}^0\pa\pi^-)
-a(1+c_V)(\pi^0\pa K^-+\sqrt{2}\pi^-\pa \bar{K}^0)\right]\non \\
&+&{\cal L}^{\rm int}(K^*,K,\pi^8,\eta_0)
\label{BKY2}
\eea
where
\bea\non
{\cal L}^{\rm int}&&(K^*,K,\pi^8,\eta_0)\\
={ig}{K}^{*0}&&
\left[\frac{1}{4\sqrt{3}}\left\{(3a+4c_A-ac_V)\pi^8\pa\bar{K}^0
-(3(a+c_A)+2(c_A^2-ac_V^2))\bar{K}^0\pa\pi^8\right\}\right.\non\\
\,\,\,\,\,\,\,\,\,\,\,\,\,\,\,\,\,\,\,\,\,\,\,\,&&\left.
+\frac{1}{2\sqrt{6}}\left\{
(c_A^2-ac_V^2)\bar{K}^0\pa\eta_0
-2(c_A-ac_V)\eta_0\pa\bar{K}^0)
\right\}\right]\non \\
-ig\bar{K}^{*0}&&
\left[\frac{1}{4\sqrt{3}}\left\{(3a+4c_A-ac_V)\pi^8\pa{K}^0
-(3(a+c_A)+2(c_A^2-ac_V^2)){K}^0\pa\pi^8\right\}\right.\non\\
\,\,\,\,\,\,\,\,\,\,\,\,\,\,\,\,\,\,\,\,\,\,\,\,
&&+\left.\frac{1}{2\sqrt{6}}\left\{
(c_A^2-ac_V^2){K}^0\pa\eta_0-
2(c_A-ac_V)\eta_0\pa{K}^0\right\}
\right] \non \\
+{ig}{K}^{*-}&&
\left[\frac{1}{4\sqrt{3}}\left\{(3(a+c_A)+2(c_A^2-ac_V^2))
{K}^+\pa\pi^8
-(3a+4c_A-ac_V)\pi^8\pa{K}^+\right\}\right.\non\\
\,\,\,\,\,\,\,\,\,\,\,\,\,\,\,\,\,\,\,\,\,\,\,\,
&&\left.+\frac{1}{2\sqrt{6}}\left\{2(c_A-ac_V)\eta_0\pa{K}^+-
(c_A^2-ac_V^2){K}^+\pa\eta_0\right\}
\right]\non \\
-{ig}{K}^{*+}&&
\left[\frac{1}{4\sqrt{3}}\left\{(3(a+c_A)+2(c_A^2-ac_V^2))
{K}^-\pa\pi^8
-(3a+4c_A-ac_V)\pi^8\pa{K}^-\right\}\right.\non\\
\,\,\,\,\,\,\,\,\,\,\,\,\,\,\,\,\,\,\,\,\,\,\,\,
&&\left.+\frac{1}{2\sqrt{6}}\left\{(2(c_A-ac_V)\eta_0\pa{K}^--
(c_A^2-ac_V^2){K}^-\pa\eta_0\right\}
\right]
\label{BKY3}
\eea

{The HLS Lagrangian broken following the BGP prescription of Ref.
\cite{BGP} is given by}
\bea\non
{\cal L}^{\rm BGP}&=&
\frac{1}{2}\left[(\pa\pi^0)^2+\left(1+\frac{4c_A}{3}\right)
(\pa\pi^8)^2+\left(1+\frac{2c_A}{3}\right)(\pa\eta_0)^2
-\frac{4\sqrt{2}c_A}{3}\pa\pi^8\pa\eta_0\right]\\
&+&\!\!
(1+c_A)\pa K^-\pa K^+ + (1+c_A)\pa K^0\pa \bar{K}^0
+\pa\pi^+\pa\pi^-\non \\
&+&\!\!
\frac{2ae^2f_{P}^2}{3}(1+c_V/3)A^2-aegf_{P}^2A\left[\rho^0+\frac{\omega}{3}
-(1+2c_V)\frac{\sqrt{2}}{3}\phi\right]
\non\\
&+&\!\!
ieA\left[
\left(1-\frac{a}{2}+c_A-\frac{ac_V}{3}\right)K^-
\stackrel{\leftrightarrow}{\pa}K^+
+\frac{ac_V}{3}K^0\stackrel{\leftrightarrow}{\pa}\bar{K}^0
+\left(1-\frac{a}{2}\right)\pi^-\stackrel{\leftrightarrow}{\pa}\pi^+
\right]\non\\
&+&\!\!
ag^2f_{P}^2\left[\frac{1}{2}({\rho^0}^2+\omega^2+(1+2c_V)\phi^2)
+\rho^+\rho^-+(1+c_V)(K^{*+}K^{*-}+\bar{K}^{*0}K^{*0})\right]
\non\\
&+&\!\!
\frac{iag}{4}\left[\rho^0(K^-\stackrel{\leftrightarrow}{\pa}K^++
K^0\stackrel{\leftrightarrow}{\pa}\bar{K}^0+
2\pi^-\stackrel{\leftrightarrow}{\pa}\pi^+)+
\omega(K^-\stackrel{\leftrightarrow}{\pa}K^+-
K^0\stackrel{\leftrightarrow}{\pa}\bar{K}^0)\right]
\non\\
&+&\!\!
\frac{iag}{2\sqrt{2}}\left[\rho^+(K^-\stackrel{\leftrightarrow}{\pa}K^0+
\sqrt{2}\pi^0\stackrel{\leftrightarrow}{\pa}\pi^-)
-\rho^-(K^+\stackrel{\leftrightarrow}{\pa}\bar{K}^0+
\sqrt{2}\pi^0\stackrel{\leftrightarrow}{\pa}\pi^+)
\right]
\non\\
&+&\!\!
\frac{iag}{2\sqrt{2}}(1+2c_V)\phi\left[K^+\lrpa K^-+K^0\lrpa \bar{K}^0
\right]
\non\\
&+&\!\!
\frac{iag(1+c_V)}{4}K^{*0}\left[\bar{K}^0\lrpa\pi^0
+\sqrt{2}\pi^+\lrpa K^-
+\sqrt{3}\pi^8\lrpa\bar{K}^0\right]\non\\
&+&\!\!
\frac{iag(1+c_V)}{4}\bar{K}^{*0}\left[\pi^0\lrpa{K}^0
+\sqrt{2} K^+\lrpa\pi^-
+\sqrt{3}K^0\lrpa\pi^8\right] \non
\\ \non
&+&\!\!
\frac{iag(1+c_V)}{4}{K}^{*-}\left[{K}^+\lrpa\pi^0
+\sqrt{2}K^0\lrpa\pi^+
+\sqrt{3}K^+\lrpa\pi^8
\right]
\\
&+&\!\!
\frac{iag(1+c_V)}{4}K^{*+}\left[\pi^0\lrpa K^-
+\sqrt{2}\pi^-\lrpa\bar{K}^0
+\sqrt{3}\pi^8\lrpa K^-\right].
\label{BGP2}
\eea
The HLS Lagrangian broken as proposed following our new scheme,
Eq.~(\ref{ours}), is given by
\bea\non
{\cal L}_{\rm new}&=&(1+c_A)\pa K^-\pa K^++(1+c_A)
\pa K^0\pa \bar{K}^0
+\pa\pi^+\pa\pi^-+(1/2)\pa\pi^0\pa\pi^0\\
\non&+&\!\!
\frac{1}{2}\left[\left(1+\frac{4}{3}c_A+
\frac{2}{3}c_A^2\right)\pa\pi^8\pa\pi^8
+\left(1+\frac{2}{3}c_A+\frac{1}{3}c_A^2\right)\pa\eta_0\pa\eta_0\right]\\
\non&-&\!\!
\frac{\sqrt{2c_A}}{3}(2+c_A)\pa\eta_0\pa\pi^8
\\
\non&+&\!\!
ae^2f_{P}^2A^2\left[\frac{2}{3}+\frac{2c_V}{9}+\frac{c_V^2}{9}\right]
-aegf_{P}^2A\left[
\rho^0+\frac{1}{3}\omega-(1+c_V)^2\frac{\sqrt{2}}{3}\phi\right]
\\
\non&+&\!\!
{ie}A\left[(1-a/2+c_A-ac_V(2+c_V)/6)K^-\stackrel{\leftrightarrow}{\pa}K^+
+\frac{ac_V(2+c_V)}{6}K^0\stackrel{\leftrightarrow}{\pa}\bar{K}^0\right.\non
\\ \non
&&\,\,\,\,\,\,\,\,\,\,\,\,\,\,\,\,\,\,\,\,\,\,\,\,\,\,\,\,\,\,\,
\left.
+(1-a/2)\pi^-\stackrel{\leftrightarrow}{\pa}\pi^+\right]
\\
\non&+&\!\!
\frac{1}{2}af_{P}^2g^2\left[{\rho^0}^2+\omega^2+(1+c_V)^2\phi^2
+2\rho^+\rho^-+2(1+c_V)(K^{*+}K^{*-}+\bar{K}^{*0}K^{*0})\right]\\
&+&\!\!
\frac{iag}{4}\left[\rho^0(K^-\stackrel{\leftrightarrow}{\pa}K^++
K^0\stackrel{\leftrightarrow}{\pa}\bar{K}^0+
2\pi^-\stackrel{\leftrightarrow}{\pa}\pi^+)+
\omega(K^-\stackrel{\leftrightarrow}{\pa}K^+-
K^0\stackrel{\leftrightarrow}{\pa}\bar{K}^0)\right]
\non\\
&+&\!\!
\frac{iag}{2\sqrt{2}}\left[\rho^+(K^-\stackrel{\leftrightarrow}{\pa}K^0+
\sqrt{2}\pi^0\stackrel{\leftrightarrow}{\pa}\pi^-)
-\rho^-(K^+\stackrel{\leftrightarrow}{\pa}\bar{K}^0+
\sqrt{2}\pi^0\stackrel{\leftrightarrow}{\pa}\pi^+)
\right]
\non\\
&+&\!\! \non
\frac{iag(1+c_V)^2}{2\sqrt{2}}\phi
\left[K^+\lrpa K^-+K^0\lrpa\bar{K}^0\right]
\\ \non
&+&\!\!
\frac{iag(1+c_V)}{4}K^{*0}\left[\bar{K}^0\lrpa\pi^0
+\sqrt{2}\pi^+\lrpa K^-
+\sqrt{3}\pi^8\lrpa\bar{K}^0\right]\non\\
&+&\!\!
\frac{iag(1+c_V)}{4}\bar{K}^{*0}\left[\pi^0\lrpa{K}^0
+\sqrt{2} K^+\lrpa\pi^-
+\sqrt{3}K^0\lrpa\pi^8\right] \non
\\ \non
&+&\!\!
\frac{iag(1+c_V)}{4}{K}^{*-}\left[{K}^+\lrpa\pi^0
+\sqrt{2}K^0\lrpa\pi^+
+\sqrt{3}K^+\lrpa\pi^8
\right]
\\
&+&\!\!
\frac{iag(1+c_V)}{4}K^{*+}\left[\pi^0\lrpa K^-
+\sqrt{2}\pi^-\lrpa\bar{K}^0
+\sqrt{3}\pi^8\lrpa K^-\right].\label{ourL}
\eea
\end{document}